\documentclass[multphys,vecphys]{svmult}

\usepackage{makeidx}         
\usepackage{graphicx}        
\usepackage{multicol}        
\usepackage[bottom]{footmisc}

\makeindex             

\begin{document}

\title*{Dwarf galaxies in the Antlia Cluster: First results}
\titlerunning{Dwarf galaxies in Antlia}
\author{A. V. Smith Castelli\inst{1,2}, L. P. Bassino \inst{1,2}, S. A. Cellone \inst{1,2}, T. Richtler \inst{3},
B. Dirsch \inst{3}, L. Infante \inst{4}, C. Aruta \inst{4} \and M. G\'omez \inst{3}
}
\authorrunning{Smith Castelli et al.}
\institute{Facultad de Ciencias Astron\'omicas y Geof\'{\i}sicas, Universidad Nacional de La Plata, Paseo del Bosque s/n, La Plata (1900), Argentina
\texttt{asmith@fcaglp.unlp.edu.ar}
\and{IALP - Conicet, Argentina
\and{Universidad de Concepci\'on, Casilla 160-C, Concepci\'on, Chile (contact address for B.D.) 
\and{Pontificia Universidad Cat\'olica de Chile, Casilla 306, Santiago 22, Chile
}}}}

\maketitle

\section{Introduction}
\label{sec:1}

The Antlia cluster (l=273$^\circ$, b=19$^\circ$) is the third nearest galaxy cluster (d=35 Mpc) after Virgo and Fornax.
In spite of its proximity, it has been poorly investigated. Its population is dominated by early type galaxies, with
dwarf ellipticals being the most abundant galaxy type \cite{FS90}.

Here we present the first results of a project aimed to study the galaxy population of the Antlia cluster. Our results correspond to the identification and classification of dwarf galaxies in the central cluster region,
extending the list of \cite{FS90} (FS90 catalogue), a photographic survey that is complete only up to
$B_\mathrm T \simeq 18$ mag ($M_\mathrm B \simeq -14.7$ mag at the Antlia cluster distance). 

The final aim of our project is to study the luminosity function, 
morphology and structural parameters of dwarf galaxies in the Antlia
cluster with a more complete sample. We also intend to investigate 
the kinematics of the cluster (50 spectra have been already obtained).

\section{Observations}
\label{sec:2}

The observations for the Antlia project were performed with the MOSAIC
camera (8 CCDs mosaic imager) mounted at the prime focus of the 4-m 
Blanco telescope at the Cerro Tololo Inter-American Observatory (CTIO).
The field of view is $36' \times 36'$.
Kron-Cousins \textit R and Washington \textit C 
filters were selected due to their known good metallicity resolution \cite{Dirsch}. 
Images from a field located at the cluster center were obtained during
April 2002, and three adjacent fields during 
March 2004. These fields cover 1.5 square degrees of the Antlia cluster.
The results presented here refer to the central field.

\section{First results}
\label{sec:3}

The detection of extended sources on the image was performed with SExtractor \cite{Sex}. In addition to 70 galaxies from \cite{FS90}, we detected many new 
dwarf galaxy candidates (dwarf ellipticals, dwarf spheroidals, and 
irregulars). The faintest candidates have $B_\mathrm T \simeq 23.4$ mag ($M_\mathrm B \simeq -9.3$ mag at the Antlia cluster distance). 
 
We find that those objects identified by SExtractor that were
considered as \textit{definite members} of 
Antlia cluster by \cite{FS90}, 
follow a well defined relation in the color-magnitude diagram ($\sigma_{color} \sim 0.08$). Also
those objects considered as \textit{likely members} by \cite{FS90}, 
are mostly located within a region of 3$\sigma$ level dispersion 
from the relation followed by \textit{definite members} (hereafter, 3$\sigma$
region). Those
objects considered as Antlia cluster \textit{possible members} 
by \cite{FS90}, tend to lie outside this region. 

Until now we have studied 17 objects, four of which are new
dwarf candidates and 13 are objects from \cite{FS90}. Only four of these 
objects (three from FS90 and one new) depart from 
the 3$\sigma$ region. The morphologies of all objects
located within the region are consistent with dwarf galaxies.

No clear color gradients are 
detected for those FS90 objects located into the 3$\sigma$ region, 
except for two galaxies (FS90 241 and FS90 188) which seem to become redder towards their centers. This, however, could in part be an effect of the different
seeing in the \textit C and \textit R exposures.

The new candidates located within the 3$\sigma$ region show clearly exponential
brightness profiles and none of them seems to posess a nucleus. 
According to the definition proposed by \cite{Gallagher}, these new
objects are faint enough to be considered as \textit{dwarf spheroidals}
($M_\mathrm B \ge -12.8$).

From its color profile, we are able to confirm that one object classified as 
Blue Compact Dwarf (BCD) by \cite{FS90} (FS90 75) is indeed a BCD galaxy: 
its color gets redder 
outwards (see e.g. \cite{BCD}) besides being the bluest object in our sample. 

Only three objects of our sample (FS90 68, FS90 72 and FS90 231) are confirmed members of the Antlia cluster through their radial velocities, obtained 
from the 6dF catalogue \cite{6dF} ($\langle V_{\rm rad}\rangle_{Antlia}\sim 2900$ km s$^{-1}$).



\printindex
\end{document}